\documentclass[twocolumn,aps,pra,amsmath,amssymb,showpacs,superscriptaddress]{revtex4}

\usepackage{ucs}
\usepackage[utf8]{inputenc} 

\usepackage{graphicx}
\usepackage{graphics}
\usepackage{latexsym}
\usepackage{amsmath}
\usepackage{amsfonts}
\usepackage{amssymb}

\begin{document}

\title{
Bell tests with photon-entanglement:\\
LHV models and critical efficiencies at the light of Wigner-PDC optics
}

\author{David Rodr{\'\i}guez}
\affiliation{Departamento de F{\'\i}sica Aplicada III, Universidad de
Sevilla, E-41012 Sevilla, Spain}
\email{drodriguez@us.es}

\date{\today}


\begin{abstract}
Within the Wigner-PDC picture of photon entanglement, detection ``errors'' are
not independent (though they may look, on average), nor can they be controlled
by means of a technological improvement on the detectors.
Those two elements make possible the interpretation of experimental evidence
without the need to exclude local realism:
for that reason, we propose the abandonment of the usual (photon, particle-based)
description of (PDC-generated) light states, in favour of an also quantum, but
field-theoretical description (QED), a description that finds a one-to-one
equivalent in that Wigner-PDC approach.
\end{abstract}


\maketitle

In recent posts we spoke of (detection probability) \emph{enhancement} and
detector inefficiencies as ``natural features of the Wigner-PDC picture of
photon entanglement''.
We are the first to admit that, left at that stage, such an assertion
looked rather vague.
Well, here we will try to give it some materiality, at least in relation to
detector inefficiencies, we have talked about enhancement somewhere else:
for instance see sidenotes of \cite{DR_WignerPDC}.
First of all, we recall that the Wigner-PDC picture \cite{WignerPDC} provides
a formalism that corresponds, one-to-one (and therefore without leaving the
orthodoxy of Quantum Electrodynamics at any stage, a fact not always
sufficiently acknowledged), to a field theoretical formulation of photon
entanglement, when generated using the well known (and nowadays almost
standardized) technique of Parametric Down Conversion (PDC).
In general we will refer to polarization-entanglement, but so far to my knowledge
such entanglement can also be explained in the same way, always within the context
of PDC, for other degree of freedom: time, path, energy... see \cite{deg_freedom}.

Recently we have proven that such Wigner-PDC picture can be cast, by means of an
elementary mathematical manipulation, into an entirely local realistic form,
where single and double detection probabilities, respectively, acquire the
expressions (following our general conventions in \cite{DR_WignerPDC} but for
minor details, for simplicity):
\begin{eqnarray}
P_{i}^{(W)}(det) & = &  \int_{\alpha} f_i(\alpha)\ W(\alpha)\ d\alpha,
\label{WPDC_eq_marginal}\\
P_{i,j}^{(W)}(det)
& = &  \int_{\alpha} \Gamma_{i,j}(\alpha) \ W(\alpha)\ d\alpha,
\label{WPDC_eq_joint}
\end{eqnarray}
where $\alpha$ is a vector of hidden variables grouping all (relevant) vacuum mode
amplitudes going into the crystal (the source), its distribution governed by
$W(\alpha)$, the Wigner function of the vacuum \cite{W_alpha}, and perhaps (if
other devices are interposed in the light rays path), also the ones entering
every other device in the setup, i.e.
\begin{eqnarray}
\alpha \equiv \alpha_s \oplus \alpha_1 \oplus \ldots \oplus \alpha_N,
\end{eqnarray}
where subindexes indicate the source (s) and the other $N$ devices in the setup.
All $\alpha$'s are defined there to be independent from each other, therefore
their joint density function factorable. In consistency, we can interpret
\begin{eqnarray}
P_{i}^{(W)}(det|\alpha) & \equiv &  f_i(\alpha), \\
P_{i,j}^{(W)}(det|\alpha) & \equiv & \Gamma_{i,j}(\alpha),
\end{eqnarray}
where detector subindexes can be omitted if all of them are physically identical.
The fact that $\Gamma_{i,j}(\alpha)$ is in general non-factorable \cite{non_fac_1},
i.e., in general
\begin{eqnarray}
\Gamma_{i,j}(\alpha) \neq f_i(\alpha) \cdot f_j(\alpha),
\end{eqnarray}
can be interpreted, using some customary terminology within the field of Quantum
Information, as ``the errors are not independent'', which does not mean that their
overall integral with respect to $\alpha$
cannot well satisfy
\begin{eqnarray}
P_{i,j}^{(W)}(det) & = &  P_{i}^{(W)}(det) \cdot P_{j}^{(W)}(det),
\end{eqnarray}
i.e.,
\begin{eqnarray}
&&\int \Gamma_{i,j}(\alpha) \ W(\alpha)\ d\alpha = \nonumber\\
&&\quad\ \
\left[ \int f_i(\alpha)\ W(\alpha)\ d\alpha \right] \cdot
\left[ \int f_j(\alpha^{\prime})\ W(\alpha^{\prime})\ d\alpha^{\prime}
\right]. \nonumber\\
\end{eqnarray}
Indeed, it is in this last sense that ``error independence'' is introduced as a
necessary hypothesis in most works on LHV \cite{LHVs}.
Now, if we want to include some technologically controlled element, an
additional dependence in the form of a ``detection efficiency'', what we
have to do is to redefine the overall detection probabilities as
\begin{eqnarray}
P_{i}^{(exp)}(det)   \equiv \mu   \cdot P_{i}^{(W)}(det),   \label{eta_marginal}\\
P_{i,j}^{(exp)}(det) \equiv \mu^2 \cdot P_{i,j}^{(W)}(det), \label{eta_joint}
\end{eqnarray}
where $0 \leq \mu \leq 1$ plays the role of such efficiency, and where
we are obviously implementing the hypothesis of ``error independence'':
this time not only ``on average'' but also for each particular realization of
$\alpha$, i.e., we also have
\begin{eqnarray}
P_{i}^{(exp)}(det|\alpha)   & \equiv & \mu   \cdot P_{i}^{(W)}(det|\alpha),  \\
P_{i,j}^{(exp)}(det|\alpha) & \equiv & \mu^2 \cdot P_{i,j}^{(W)}(det|\alpha).
\end{eqnarray}
For a Bell experiment, nevertheless, it is only the average probabilities
(integrated in $\alpha$) that present interest, because they are the ones
used to evaluate the overall result of the test.
Such probabilities are estimated from the number of counts registered on
a certain time-window $\Delta T$ (sufficiently large);
for instance, we can estimate
\begin{eqnarray}
P_{i}^{(exp)}(det) \approx \frac
{ n.\ joint \ det.\ (i,j) \ in\ \Delta T }
{ n.\ marg. \ det.\ (j)   \ in\ \Delta T}, \label{estimation_marg}
\end{eqnarray}
and finally we should have, assuming, as we have seen, independence of errors
``on average'': 
\begin{eqnarray}
P_{i,j}^{(exp)}(det) \approx P_{i}^{(exp)}(det) \cdot P_{j}^{(exp)}(det).
\label{estimation_joint}
\end{eqnarray}

At this point, it is time to recall that a violation of a Bell inequality
only happens when the marginal detection probability (let us for
simplicity assume that is equal for all detectors) surpasses a certain
threshold, customarily known as ``critical efficiency'' (which at least as
far as this framework is concerned, is a clearly misleading name: it should
be substituted by, for instance, that of ``critical detection probability'').

However,  the Wigner-PDC picture can be formulated, as proven in \cite{DR_WignerPDC},
as a local theory, therefore abiding to all possible Bell inequalities, even for
$\mu = 1$.
For other values of $\mu$, marginal probabilities cannot be increased (nor
can the joint ones), so therefore there is no physical way of externally raising
those ``efficiencies'' so as to violate the inequality.
In other words, unless the events of ``failed detection'' and ``absence of emission''
can be somehow distinguished (which seems difficult), the customary game of
Local Hidden Variable (LHV) models and critical efficiencies, well defined as it
may be, has never found (to my knowledge) an interpretation as realistic as the one
we obtain with the Wigner-PDC picture that we advocate here: critical efficiencies 
are simply bounds on the detection rates that we can physically obtain.

Just as a proposal, we think it is not at this point unjustified to suggest
the abandonment of the usual (photon, particle-based) Hilbert space description
of all (PDC-generated) light states, in favour of an also quantum,
but field-based description (QED), a description that finds a one-to-one
equivalent in a picture of continuous, random variables (the Wigner-PDC picture),
compatible with local realism and that gives rise, as natural features, to
(detection probability) \emph{enhancement} and (detection) \emph{inefficiency},
this last another misleading term: we should perhaps use something like
``absence of detection'' or simply ``reduced detection probability''.
While the first of those two phenomena is simply ignored in the customary
description, the second is introduced as a mere external postulate to match
observations.

\emph{Last comments:}
all we are doing in this paper is \emph{comparing} two different models;
of course, other more refined ones can also be build, introducing
non-punctual detectors and many other features.
All this modifications are desirable but not of our interest here: we just
wanted to show that there is a basic feature (reduced detection efficiency)
arising naturally from the mathematical structure of one of them, while it is
absent in the other...
a feature that seems to fit well with the experimental evidence (or absence
of it, given the elusive character, after several decades, of a conclusive
proof of the incompatibility of QM and local realism: so far, all observed
violations of a Bell inequality do always involve additional hypothesis
\cite{ad_hypothesis}).




\begin{thebibliography}{99}
\bibitem{PDC_explanation}
Parametric Down Conversion (PDC): a pair of entangled photons is obtaining by pumping a
laser beam into a non-linear crystal.
Within the Winger picture \cite{WignerPDC}, the pair generation can be interpreted
as the non-linear mix of the laser frequency with the vacuum components.
\bibitem{DR_WignerPDC}
D. Rodr{\'i}guez,
``Wigner-PDC description of photon entanglement can still be made
completely local''.
ArXiv.
\bibitem{non_fac_1}
This non-factorability arises naturally in the LHV's cited in \cite{LHVs}, when one
tries to introduce the ``non-detection'' instructions as a purely random element,
i.e., we do not distinguish between states $\lambda$ predetermining or not such
non-detections.\\
Besides, we are aware, following our recent work in \cite{DR_fac}, that non-factorability
can only be the result of other hidden variables (HV), relevant for the state of the
source, not appearing explictly in the expressions for the predictions of the
theory: this makes perfect sense, as here we are only setting the focus on vaccum
amplitude as HV's... what about the laser, the crystal?
\end{thebibliography}
\end{document}